\documentclass[fleqn,12pt]{wlscirep}
\usepackage[utf8]{inputenc}
\usepackage[T1]{fontenc}
\usepackage{soul}
\def\pig{\pi_{\rm{gas}}}
\def\pil{\pi_{\rm{liq}}}
\def\rhog{\rho_{\rm{gas}}}
\def\rhol{\rho_{\rm{liq}}}
\def\rhob{\rho_{\rm{bulk}}}
\title{Universality, scaling and collapse in supercritical fluids}

\author[1]{Min Young Ha}
\author[1]{Tae Jun Yoon}
\author[2,3,*]{Tsvi Tlusty}
\author[4,*]{YongSeok Jho}
\author[1,*]{Won Bo Lee}

\affil[1]{School of Chemical and Biological Engineering, Institute of Chemical Processes, Seoul National University, Seoul 08826, Republic of Korea}
\affil[2]{Center for Soft and Living Matter, Institute for Basic Science (IBS), Ulsan 44919, Republic of Korea}
\affil[3]{Department of Physics, Ulsan National Institute of Science and Technology (UNIST), Ulsan 44919, Republic of Korea}
\affil[4]{Department of Physics and Research Institute of Natural Science, Gyeongsang National University, Jinju 52828, Republic of Korea}

\affil[*]{e-mail: tsvitlusty@gmail.com, ysjho@gnu.ac.kr, wblee@snu.ac.kr}

\begin{abstract}

\end{abstract}
\begin{document}

\maketitle

\section*{Introduction}
\textbf{
    The Supercritical Fluid (SCF) is known to exhibit  salient dynamic and thermodynamic crossovers~\cite{xu2005relation,simeoni2010widom,mcmillan2010fluid} and inhomogeneous molecular distribution~\cite{tucker1999solvent,nishikawa2000inhomogeneity,yoon2017monte,yoon2017molecular}. But the question as to what basic physics underlies these microscopic and macroscopic anomalies remains open. Here, using an order parameter extracted by machine learning, the fraction of gas-like (or liquid-like) molecules, we find simplicity and universality in SCF: First, all  isotherms of a given fluid collapse onto a single master curve described by a scaling relation. The observed power law holds from the high-temperature and pressure regime down to the critical point where it diverges. Second, phase diagrams of different compounds collapse onto their master curves by the same scaling exponent, thereby demonstrating a putative law of corresponding supercritical states in simple fluids. The reported results support a model of the SCF as a mixture of two interchangeable microstates, whose spatiotemporal dynamics gives rise to unique macroscopic properties.
}

\section*{Main text}

Almost two centuries after its discovery by Cagniard de la Tour \cite{CagniardDeLaTour1822}, understanding the supercritical fluid (SCF) remains a challenge. This exotic state of matter exhibits an anomalous blend of liquid-like and gas-like traits that is beyond the scope of conventional fluid theories~\cite{mcmillan2010fluid,bolmatov2013thermodynamic}. This distinctive nature also renders the SCF a promising candidate for potential industrial applications~\cite{eckert1996supercritical,savage1995reactions}. SCF simultaneously manifests microscopic inhomogeneities and macroscopic crossover phenomena. Therefore, developing a theoretical framework that can capture the physics of SCF at all the relevant scales is still an open problem. Inspired by recent applications in physics~\cite{carrasquilla2017machine,van2017learning,cubuk2015identifying,schoenholz2016structural}, we used machine learning classifiers to identify distinct liquid-like and gas-like microstructures coexisting in SCF~\cite{ha2018widom}. Linking supercritical anomalies to fluctuations between the microstates, we proposed the number fraction of gas-like molecules, $\pig$ , as a useful order parameter of SCF.

In this Letter, we report on universal scaling of the supercritical phase diagram obtained using the order parameter $\pig$. Unlike standard scaling laws, which are usually restricted to the vicinity of the critical point, the reported scaling relation spans through the whole ``Widom delta"\cite{ha2018widom}, from the critical point up to high temperature and pressure. The divergence of the scaling functions accurately pinpoints the liquid-gas critical point (LGCP). Furthermore, the scaling exponents are found to be universal among three simple fluids we examined, Ar, $\rm{CO_{2}}$ and $\rm{H_{2}O}$. These results provide evidence that the physics of SCF is governed by the scaling relation of microscopic inhomogeneity.

\begin{figure}
        \includegraphics[width=16.0cm]{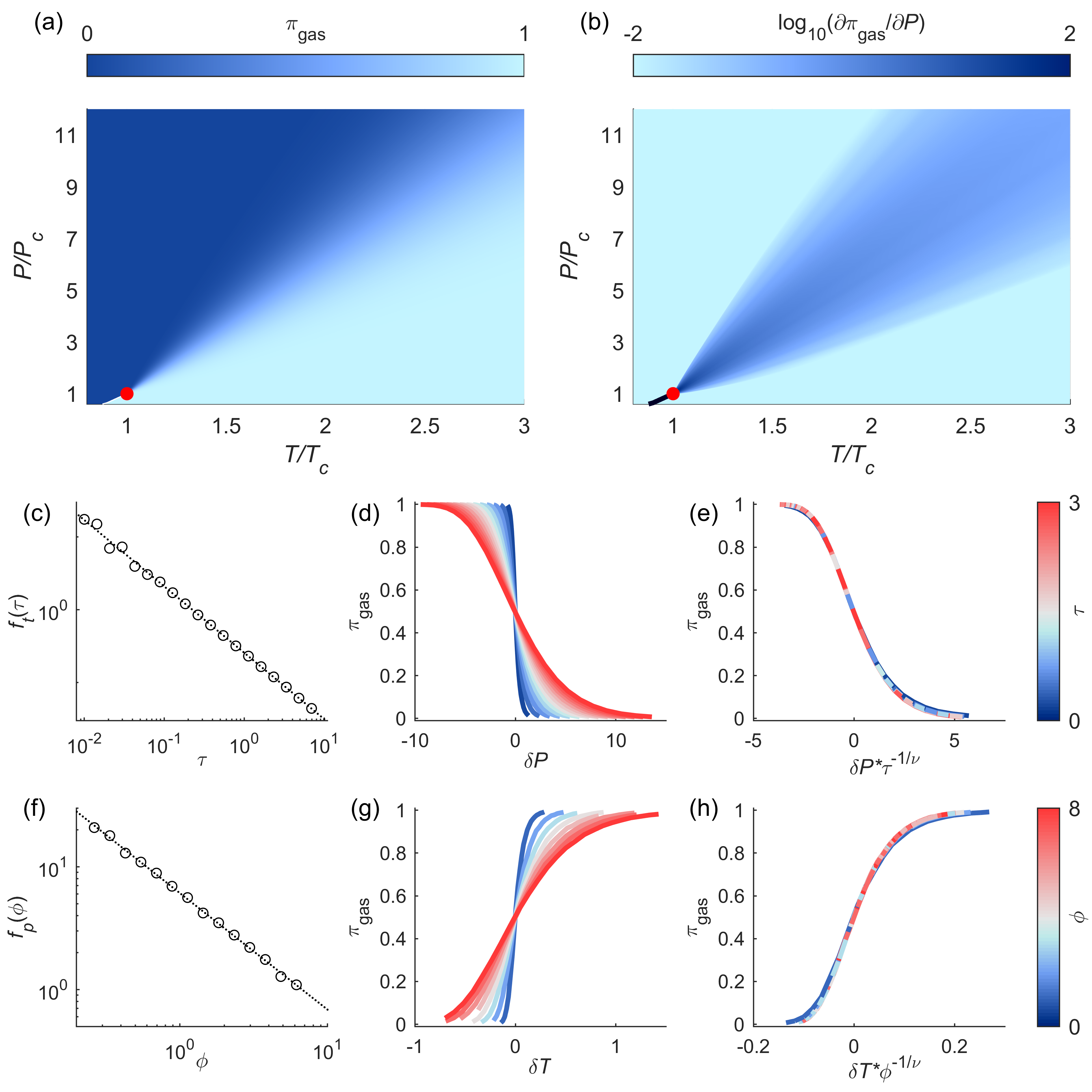}
    \centering
    \caption{
        \textbf{Phase diagram and scaling analysis of the order parameter $\pig$.}
        \textbf{(a)} Phase diagram of $\pig$. Below the liquid-gas critical point (LGCP, red dot), $\pig$ jumps discontinuously across the vapor-liquid equilibrium (VLE) line. Beyond the LGCP, $\pig$ varies continuously across the ``Widom delta".  \textbf{(b)} Phase diagram of the pressure derivative of $\pig$. Note that the map is expressed in log-scale, with range restricted to four orders of magnitude. The derivative diverges at the LGCP, and smears out into the deeply supercritical regime.
        \textbf{(c)} The isothermal scaling function $f_t(\tau)$ (see text), for $1.01T_c<T<10.0T_c$.
        \textbf{(d-e)} Raw (d) and rescaled (e) curves of $\pig$ along different isotherms, $1.2T_c<T<3.8T_c$. The color bar on the right end denotes the temperatures.
        \textbf{(f)} The isobaric scaling function $f_p(\phi)$, for  $1.26P_c<P<6.16P_c$.
        \textbf{(g-h)} Raw (g) and rescaled (h) curves of $\pig$ along different isobars, $2.0P_c<P<9.5P_c$. The color bar on the right end denotes the pressures.
    }
    \label{fig:Scaling}
\end{figure}

Motivated by investigations of supercritical scaling in systems with Ising~\cite{anisimov1995nature} and quantum criticality~\cite{terletska2011quantum,furukawa2015quantum}, we examined the variation of $\pig$ in the phase diagram (Fig.~\ref{fig:Scaling}). Before we expound on the results, we briefly explain the main features of our approach. The computational method we use was developed and tested in a previous work~\cite{ha2018widom} (see Methods and Supplementary Information): Neural networks label individual particles in SCF as liquid-like and gas-like, based on their local structures. For a given temperature and pressure, $\pig$ is the ensemble-averaged number fraction of gas-like particles. 

As a reference line, we use the supercritical gas-liquid boundary (SGLB), defined as the loci of equal liquid and gas fractions, $\pig = \pil = 0.5,$ in the supercritical domain $(T>T_c)$. Taking the SGLB, which is also the line of maximal instability, as a reference is based on two observations: (i) near the LGCP, the Widom line and the SGLB closely coincide~\cite{ha2018widom}, and (ii) the vapor-liquid equilibrium line (VLE), the LGCP, and the Widom line can be framed together within an extended corresponding states principle~\cite{banuti2017similarity}. Therefore, we employ the postulate that the SGLB is the extension of the VLE and LGCP into the supercritical region of the phase diagram. 

The phase diagram of the order parameter $\pig$, as computed by the machine learning classifiers (Fig. \ref{fig:Scaling}(a)), exhibits discontinuity across the VLE line. In contrast, beyond the LGCP, $\pig$ is continuous throughout the ``Widom delta" -- the deltoid liquid-gas coexistence region emanating from the LGCP. The diagram of the pressure derivative, $\partial \pig / \partial P$, highlights the divergence at the LGCP (Fig. \ref{fig:Scaling} (b)). 

It is instructive to define the functions $f_t$ and $f_p$, the isothermal and isobaric gradients of $\pig$ at the SGLB:
\begin{equation}
        f_t(T) = \frac{\partial\pig}{\partial P}\Bigr|_{\rm{SGLB}}, \;
    f_p(P) = \frac{\partial\pig}{\partial T}\Bigr|_{\rm{SGLB}}.
    \label{eqn:ftfp}
\end{equation}
Following the standard scaling procedure, we compute the reduced temperature  and pressure with respect to the critical point, i.e. $\tau=(T-T_c)/T_c$ and $\phi=(P-P_c)/P_c$. In these reduced coordinates, the functions  $f_t(\tau)$ (Fig.~\ref{fig:Scaling}(c)), and $f_p(\phi)$ (Fig.~\ref{fig:Scaling}(f)) nicely fit to power-laws:

\begin{equation}
        f_t(\tau) = A_t\tau^{-1/\nu_t},
    f_p(\phi) = A_p\phi^{-1/\nu_p}~.
    \label{eqn:power}
\end{equation}

\noindent The prefactors $A_t$, $A_p$ and the exponents, $\nu_t$ and $\nu_p$, were extracted by fitting the data to equation (\ref{eqn:power});
 the estimated exponents are $\nu_t=1.10$ and $\nu_p=1.05$.

The second postulate we assume and examine is that different isothermal curves of $\pig$ are all \textit{isomorphic}, with scales determined by the scaling function $f_t$. Fig.~\ref{fig:Scaling}(d) shows isotherms of $\pig$ as a function of $\delta P(T) = P - P_{\rm{SGLB}}(T)$, the pressure shift from the SGLB. Rescaling $\delta P$ by $\tau^{-1/\nu_t}$, we observe collapse into a single master curve (Fig.~\ref{fig:Scaling}(e)), which captures  the whole supercritical phase diagram. Similarly, different isobars of $\pig$ as a function of the temperature shift $\delta T(P) = T - T_{\rm{SGLB}}(P)$ also collapse  into a single curve by rescaling $\delta T$ by $\phi^{-1/\nu_p}$ (Fig.~\ref{fig:Scaling}(f-h)).

The scaling functions  $f_t$ and $f_p$ diverge at the critical point, as evident from their power law dependence (Fig.~\ref{fig:Scaling}(c) and (f)).  So far, however, the critical values,  $T_c$ and $P_c$, used for obtaining the power laws (equation (\ref{eqn:power})) were not fitting parameters, but rather pre-determined (Supplementary Information). To test if our model is able to capture the critical point itself, we tried a more flexible form of $f_t(T)$ and $f_p(P)$:

\begin{align}
        f_t(T) = A_t\left(\frac{T - T_c^*}{T_c^*}\right)^{-1/\nu_t},\;
    f_p(P) = A_p\left(\frac{P - P_c^*}{P_c^*}\right)^{-1/\nu_p}~,
    \label{eqn:CP}
\end{align}
where  $T_c^*$ and $P_c^*$ are  free fitting parameters. The numerical fitting yielded $(T_c^*, P_c^*) = (1.3087, 0.1374)$, comparable to values determined from the law of rectilinear diameters, $(1.3047, 0.1202)$, or from van der Waals hypothesis, $(1.3449, 0.1428)$. Interestingly, the values  are in the vicinity of the critical point of full LJ potentials~\cite{caillol1998critical,potoff1998critical}, $(1.326, 0.1279)$.

It is worth noting that the liquid/gas classifiers were trained with information only  on the local structure of individual particles. Considering the strong volume fluctuations in the supercritical state, the classifiers are blind to the macroscopic state of the system, which is only loosely encoded in the ensemble average of $\pig$. Nevertheless, the reported results put forth the number fraction of the two particle types as a measure that  captures the critical behavior and the bulk characteristics of the system, which are collectively determined  by spatiotemporal dynamics of the microstructure.

\begin{figure}
        \includegraphics[width=17.2cm]{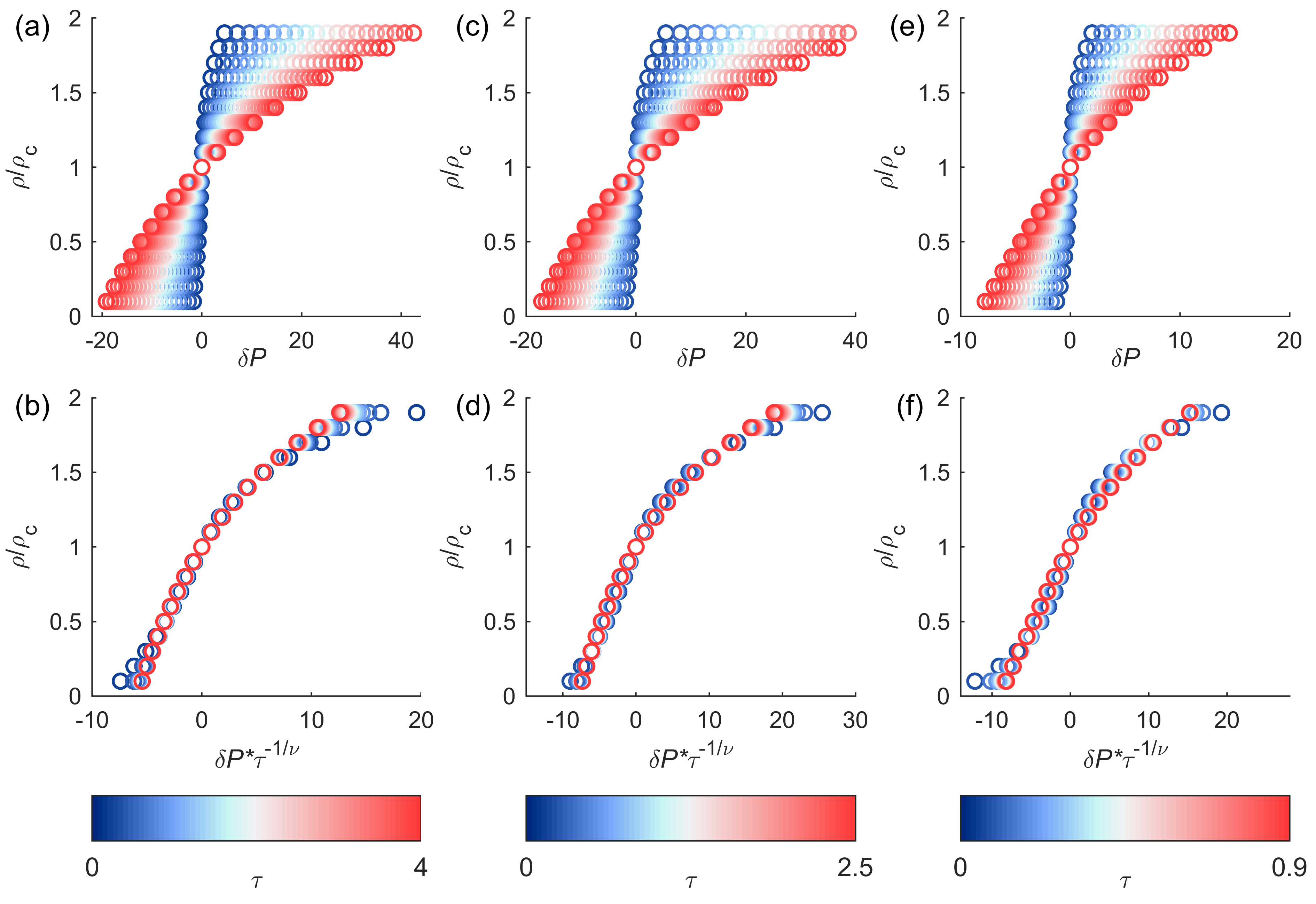}
    \centering
    \caption{\textbf{Density rescaling in the supercritical region.}
    Density rescaling by the scaling exponent $\nu_t$, applied to (a-b) LJ fluid, (c-d) carbon dioxide and (e-f) water. The data of $\rm{CO_2}$ and $\rm{H_2O}$ were acquired from the NIST database~\cite{nist}. In this figure, the density difference was defined as $\delta P = P - P_{\rm{isochore}}$, where $P_{\rm{isochore}}$ is defined as the pressure at the critical isochore, i.e. $\rho = \rho_c$.
    }
    \label{fig:Density}
\end{figure}

At this point, one might naturally question the relation between $\pig$ and experimental observables, e.g. the bulk density $(\rhob=N/V)$, since $\pig$ is only accessible via molecular simulations~\cite{yoon2018probabilistic}. To resolve this issue, we define the density of gas-like and liquid-like states as $\rhog=N_{\rm{gas}}/V_{\rm{gas}}$ and $\rhol=N_{\rm{liq}}/V_{\rm{liq}}$. The data exhibits several `empirical' relations between $\rhog$, $\rhol$ and $\rhob$ (see Figs. S3-S4 and comments there). First, the bulk density equals the weighted sum of the liquid-like and gas-like densities:

\begin{equation}
        \rhob = \rhol \pil + \rhog \pig.
    \label{eqn:rhobulk}
\end{equation}

\noindent Note that equation (\ref{eqn:rhobulk}) is not merely the outcome of volume and number conservation, which lead to $\rhob = \pig / \rhog + \pil / \rhol$. Denoting liquid-like and gas-like densities at SGLB as $\rhol^{\rm{o}}$ and $\rhog^{\rm{o}}$, and defining the isothermal deviation from the SGLB as $\Delta\rhol(T,P) = \rhol(T,P) - \rhol^{\rm{o}}(T)$ and $\Delta\rhog(T,P) = \rhog(T,P) - \rhog^{\rm{o}}(T)$, equation~(\ref{eqn:rhobulk}) can be rewritten as:

\begin{equation}
 \rhob = \frac{1}{2}\left(\rhol+\rhog\right) +  \Delta\pi\left(\Delta\rhol
  - \Delta\rhog\right)  + \Delta\pi\left(\rhol^{\rm{o}} - \rhog^{\rm{o}}\right)~,
    \label{eqn:expansion}
\end{equation}

\noindent where $\Delta\pi = \pil - 0.5 = 0.5 - \pig$.

Examining the data in Figs. S3-S4, one finds that the three terms on the r.h.s. of equation (\ref{eqn:expansion})  exhibit the following dependence:\ (i) $(\rhol+\rhog)/2$ is a function of $\pig$, (ii) $\Delta\rhol-\Delta\rhog$ is also a function of $\pig$, and (iii) $\rhol^{\rm{o}}-\rhog^{\rm{o}}$ is a weak function of the temperature $T$, saturating to a finite value. Hence, if the range of $T$ is not too wide, or if $T$ is  high enough, then $\rhob$ is, approximately, a function of $\pig$.
From all this, one concludes that  the rescaling of $\delta P$ with $\tau^{-1/\nu_t}$  (Fig.~\ref{fig:Scaling}) applies also  to $\rhob$. 

This conjecture was first tested in a supercritical LJ\ fluid (Fig.~\ref{fig:Density}(a-b)). We find that, as long  the reduced temperature $\tau$ is sufficiently far from the  LGCP, it can be reduced to a single line using the same scaling exponent $\nu_t$ of the order parameter $\pig$.
This scaling is not restricted to the LJ toy model, and seems to be universal among simple fluids, at least for the three examples we examined: Fig.~\ref{fig:Density}(c-f) show the rescaling of  the phase diagrams of supercritical $\rm{CO_2}$ and $\rm{H_2O}$ (data taken  from the NIST chemistry webbook~\cite{nist}). To avoid dependence on   machine learning techniques, we defined $\delta P$ as the deviation from the critical isochore, instead of the SGLB. With the same scaling exponent $\nu_t$ acquired from LJ fluid, the bulk densities of supercritical $\rm{CO_2}$ and $\rm{H_2O}$ also collapse to a single line. Similar rescaling is obtained  for isobaric density curves (Fig. S6). Overall, these results suggest that  supercritical scaling in the Widom delta is universal in simple fluids. The observed scaling also supports the framework proposed in this Letter, of understanding the SCF as an inhomogeneous mixture of two microstates.

In conclusion, by analyzing the behavior of the order parameter $\pig$, we found material-independent scaling relations for the supercritical state of several simple fluids. The supercritical gas-liquid boundary serves as the reference line of the SCF, from which the properties of SCF can be rescaled into a master curve. The reported results put forth a  picture of the SCF as an inhomogeneous mixture of microstates that collectively determine both microscopic and macroscopic features. 

Finally, it is worth pointing that the Widom delta in Fig.~\ref{fig:Scaling}(a-b) is akin to the ``fan-shaped'' quantum critical region found near metal-insulator transitions, where universality and scaling are expected~\cite{vuvcivcevic2013finite,furukawa2015quantum,terletska2011quantum,lamsal2009search}. In the language of quantum critical phenomena, systems in deeply critical regions of the Widom delta cannot `know' whether they are  liquid or gas. Only when the system comes closer to the LGCP, it `notices' to which phase  it belongs and then bifurcates into the liquid-like and gas-like branches. Below the critical temperature, the system divided into liquid/gas states separated by first-order phase transition, analogous to the quantum ground states at zero temperature. One potential direction for future study would be testing whether the `scaling region' universally exists for both quantum and classical phenomena.

\section*{Methods}
All molecular dynamics simulations were performed using LAMMPS~\cite{plimpton1995fast}. Fluid particles interacted via a 12-6 Lennard-Jones pairwise potential, truncated at cutoff distance of $r_{\rm{cut}}=3.0\sigma$ with standard correction on long-range energy contribution. Nose-Hoover thermostat and barostat were used to maintain the temperature and pressure of the system at the desired value. Isothermal and isobaric properties were calculated in \textit{NVT} and \textit{NPT} ensemble, where the thermodynamic properties were averaged over 10,000,000 time steps. Machine learning analyses were performed on snapshots of configurations dumped every 10,000 time steps. Voro++ library was used for the Voronoi tessellation~\cite{rycroft2009voro++}.

Machine learning classifiers were implemented using deep convolutional neural networks~\cite{goodfellow2016deep}, designed following the VGGNet~\cite{simonyan2014very}. Training data were the local structure information of subcritical liquid and gas particles in saturated liquid and vapor phases, sampled by Gibbs-ensemble Monte Carlo simulation~\cite{frenkel2001understanding} at near-critical temperature of $T/T_c=0.97$. The neural network was trained by supervised learning approach, optimizing the node weights to maximize classification accuracy of liquid and gas particles. We trained 24 neural networks of same architecture with different training sets, and their training accuracies were $0.95\pm0.002$. The trained neural networks could classify saturated liquid and vapor particles at $T/T_c=0.9$ with 100\% accuracy.

\section*{Acknowledgements}
This research was supported by Creative Materials Discovery Program through the National Research Foundation of Korea (NRF) funded by Ministry of Science and ICT (NRF-2018M3D1A1058624), and NRF Grant funded by Korean Government (NRF-2017H1A2A1044355, NRF-2018R1A2B6006262).

\section*{Author contributions}
M.Y.H., Y.J. and W.B.L. designed the simulations. M.Y.H. and T.J.Y. performed the simulations and collected data. M.Y.H., T.T. and Y.J. interpreted the results. M.Y.H. wrote the manuscript with the assistance of T.T. and Y.J. W.B.L. headed this project.

\section*{Additional information}
Online Supplementary Information is available. Correspondence and requests for materials should be addressed to Y.J., T.T. or W.B.L.

\section*{Competing financial interests}
The authors declare no competing financial interests.

\bibliography{references}

\end{document}